\documentclass[superscriptaddress,onecolumn,showkeys,preprintnumbers,amsmath,amssymb]{revtex4}

\usepackage{graphicx}
\usepackage{dcolumn}
\usepackage{bm}
\usepackage{color}

\begin{document}

\title{Activation of effector immune cells promotes tumor stochastic extinction: A homotopy analysis approach}

\author{Josep Sardany\'es}
\thanks{Corresponding author: J. Sardany\'es (josep.sardanes@upf.edu).}
\affiliation{Complex Systems Lab, Parc de Recerca Biom\`edica de Barcelona, Dr. Aiguader, 88. 08003 Barcelona, Spain}
\affiliation{Institut de Biologia Evolutiva, Pg. Maritim de la Barceloneta 37, 08003 Barcelona, Spain.}

\author{Carla Rodrigues}
\affiliation{ESTS - Technology Superior School of Setubal, Department of Mathematics, Campus do IPS, Rua Vale de Chaves, Estefanilha, 2914-761 Setubal, Portugal.}

\author{Cristina Janu\'ario}
\affiliation{ISEL - Engineering Superior Institute of Lisbon, Department of Mathematics, 
Rua Conselheiro Em\'{\i}dio Navarro 1, 1949-014 Lisboa, Portugal.}

\author{Nuno Martins}
\affiliation{Center for Mathematical Analysis, Geometry and Dynamical Systems,
Mathematics Department,
Instituto Superior T\'ecnico, Universidade de Lisboa
Av. Rovisco Pais, 1049-001 Lisboa, Portugal.}

\author{Gabriel Gil-G\'omez}
\affiliation{Apoptosis Signalling Group, Cancer Research Programme, IMIM (Hospital del Mar Medical Research Institute), Dr. Aiguader, 88 08003 Barcelona, Spain}

\author{Jorge Duarte}
\affiliation{ISEL - Engineering Superior Institute of Lisbon, Department of Mathematics, 
Rua Conselheiro Em\'{\i}dio Navarro 1, 1949-014 Lisboa, Portugal.}
\affiliation{Center for Mathematical Analysis, Geometry and Dynamical Systems,
Mathematics Department,
Instituto Superior T\'ecnico, Universidade de Lisboa
Av. Rovisco Pais, 1049-001 Lisboa, Portugal.}


\begin{abstract}
In this article we provide homotopy solutions of a cancer nonlinear model describing the dynamics of tumor cells in interaction with healthy and effector immune cells. We apply a semi-analytic technique for solving strongly nonlinear systems - the \textit{Step Homotopy Analysis Method} (SHAM). This algorithm, based
on a modification of the standard homotopy analysis method (HAM), allows to obtain a one-parameter family
of explicit series solutions. By using the homotopy solutions, we first investigate the dynamical effect of the activation of the effector immune cells in the deterministic dynamics, showing that an increased activation makes the system to enter into
chaotic dynamics via a period-doubling bifurcation scenario. Then, by adding demographic stochasticity into the homotopy solutions, we 
show, as a difference from the deterministic dynamics, that an increased activation of the immune cells facilitates cancer clearance involving tumor cells extinction and healthy cells 
persistence. Our results highlight the importance of therapies activating the effector immune cells at early stages of cancer progression.
\end{abstract}


\maketitle

\section{Introduction}
Exact solutions for nonlinear equations
are difficult to obtain, and a handful of novel methods and techniques, either analytical or numerical,
have been developed. The nature of the interactions in biological systems gives place to
nonlinear dynamics that can generate, for some parameter values, very
complicated dynamics e.g. chaos. Hence, advances to better characterize the dynamics for
nonlinear systems turn out to be extremely useful to analyze and understand such systems. 
As far as analytical approaches are concerned, various perturbation
techniques are frequently applied in science and engineering, and they do
help us to enhance our understanding of nonlinear phenomena. Nevertheless,
many perturbation methods are only valid and effective for weakly nonlinear
problems, due to their strongly dependence upon small/large physical
parameters. On the other hand, the so called traditional non-perturbation
methods, such as the artificial small parameter method \cite{Lyap1992}, the $%
\delta$-expansion method \cite{Karm1990,Awr1998}, and the
Adomian's decomposition method \cite{Adomian1976,Adomian1991,Rach1984,Adomian1984a}, are
known to be formally independent of small/large physical parameters. However, all of these non-perturbation techniques are, in fact, valid for weakly
nonlinear problems and they can not ensure the convergence of solutions
series for strongly nonlinear systems. As a consequence, in recent years there has been a growing
interest in obtaining continuous solutions for nonlinear dynamical
systems by means of analytical or semi-analytical techniques. The \textit{homotopy analysis method} (HAM), initially
proposed by Liao \cite{Liao1992,Liao2003}, can be used to obtain convergent series solutions of strongly nonlinear
problems (including e.g., ordinary differential equations, partial differential
equations, algebraic equations, and differential-integral equations). 

Unlike perturbation methods, the HAM is independent of small/large physical
parameters. Differently from perturbation and non-perturbation methods, the HAM is valid even for strongly
nonlinear problems and it is characterized by two central aspects: (i) a
great freedom to choose proper linear operators and base functions to
approximate a nonlinear problem, and (ii) the use of an artificial parameter
that represents a simple way to adjust and control the convergence region
and rate of convergence of the series solution. Based on \textit{homotopy},
a fundamental concept in topology and differential geometry \cite{Sen1983},
the HAM allows us to construct a continuous mapping of an initial guess
approximation to the exact solutions of the considered equations, using a
chosen linear operator. Indeed, the method enjoys considerable freedom in
choosing auxiliary linear operators. The
HAM represents a truly significant milestone that converts a complicated nonlinear problem into an
infinite number of simpler linear sub-problems \cite{Liao2007}.
Since Liao's work \cite{Liao2003}, the HAM has been
successfully employed in the fractional Lorenz system \cite{Alomari2010}, in fluid dynamics \cite{Hayat2005}, in the Fitzhugh-Nagumo model \cite{Li2006}, as well as to obtain soliton solutions also for the Fitzhugh-Nagumo system \cite{Abbasbandy2008}. This semi-analytical technique has been also used in complex systems in ecology 
\cite{Arafa2011,Putcha2012}, in epidemiology \cite{Liao2009}, as well as in models of interactions between tumors and oncolytic viruses \cite{Usha2012}.

In the present article we apply the HAM to obtain solutions of a cancer growth model proposed by Itik and Banks \cite{Itik2010}.  Such a
model, based on Volterra-Lotka predator-prey dynamics, describes the
interactions between tumor, healthy, and effector immune cells (CD8 T cells i.e., cytotoxic lymphocytes, CTLS). Predator-prey or competition Volterra-Lotka systems are known to display deterministic chaos for systems with three or more dimensions \cite{Hastings1991,Vano2006,Gakkhar2003,Tang2002}. Together with the model by Itik and Banks, several other theoretical models have addressed the dynamics of cancer and tumor cells \cite{Kuznetsov1994,dePillis2003,Kirschner1998}. Interestingly, the model by Itik and Banks  can be considered as being qualitatively
validated with experimental data, because its parameter values were chosen
to match with some biological evidences. This model could be thus considered as being qualitatively validated with experimental data  \cite{dePillis2003,Letellier2013}. Motivated by
the characterization of chaos provided by Itik and Banks, a
collection of questions pertaining to chaotic tumor behavior in terms of
symbolic dynamics and predictability as well as to the control of healthy cells behavior corresponding to
physiological relevant parameter regions, have been recently addressed in 
Refs. \cite{Duarte2013} and \cite{Duarte2014}, respectively. In fact, chaos in tumor dynamics and its property of sensitivity to initial conditions have been suggested to have numerous
analogies to clinical evidences \cite{Denis2012a,Denis2012b}. 

Numerical algorithms have been extremely important to investigate complex dynamical systems such as cancer. However, they allow us to analyze the dynamics at
discrete points only, thereby making impossible to obtain continuous
solutions. By means of the HAM, accurate approximations allow a good semi-analytical description of the time variables, making also possible 
to use the homotopy solutions to explore the model dynamics, as well as to investigate possible scenarios of tumor clearance, either deterministic or stochastic. This is the aim that we pursue in this contribution. Specifically, we will calculate the homotopy solutions of the cancer model by means of the step homotopy analysis method (SHAM, see \cite{Alomari2010}). Then, the homotopy solutions will be used to explore the effect of a key parameter in the population dynamics: the activation of the immune system cells due to tumor antigen recognition, given by parameter $r_3$ (see next Section). As we will show, the system is very sensitive to this parameter, and its change can involve the shift from order to chaos. This key parameter is especially interesting because modulates the response of the immune system against tumor cells and, as we will show, the dynamics is especially sensitive to $r_3$. Despite its importance, the dependence of the dynamics of the model under investigation on $r_3$ remains poorly explored (see \cite{Duarte2014} for the analysis of a narrow range of $r_3$ values within the framework of chaotic crises and chaos control). Moreover, the impact of this parameter on possible extinction scenarios of tumor cells due to demographic fluctuations has, as far as we know, not being investigated. Interestingly, several therapeutic methods, that will be discussed in this article, are currently available to clinically manipulate this parameter, thus being a realistic candidate to fight against tumor progression.

Finally, we will use the homotopy solutions to investigate the role of demographic stochasticity in the dynamics of 
the model, paying special attention to the role of noise in potential scenarios of tumor clearance and persistence of healthy cells due to changes in the activation levels of effector immune cells.

\section{Cancer mathematical model}
In this article we analyze a cancer mathematical model initially studied by Itik and Banks \cite{Itik2010}. The model describes the dynamics of three interacting cell populations: tumor
cells, healthy  cells and effector immune cells i.e.,  CD8 cytotoxic T-cells, CTLs. 
Effector cells are the
relative short-lived activated cells of the immune system that defend the
body in an immune response. Similarly to previous cancer models \cite%
{Kuznetsov1994,dePillis2003,Kirschner1998,Kuznetsov2001,dePillis2006,Itik2009,Baizer1996}, this model describes the competition dynamics of these three interacting
cell types in a well-mixed system (e.g., liquid cancers such as leukemias or multiple lymphomas). Among several biologically-meaningful assumptions (see \cite{Itik2010}), the model assumes that the antitumor effect of the immune system response is carried out by cytotoxic T-cells i.e., mediated by the T-cell based adaptive arm. Alpha-beta T-cells are activated upon recognition of their cognate tumor specific antigens by the cell surface T-Cell Receptor (TCR) in the form of small peptides presented in the context of the major histocompatibility complex (MHC) molecules. CD8 T-cells are responsible for direct cell mediated cytotoxicity following activation by antigen presenting cells (APCs) and are thought to be central players in the anti-tumor immune response. 
To achieve full activation, the signal emanating from the TCR has to be enhanced by messages sent by costimulatory molecules such as CD28 also present in the surface of the T-cell. Failure of the engagement of costimulatory proteins, activation of coinhibitory receptors such as CTLA-4 or PD-1 or the presence of CD4 regulatory ($T_{reg}$) T cells may lead to the failure of the activation of the T-cell or to the downregulation of the immune response. Disarming these inhibitory mechanisms off may lead to the reactivation of the antitumor immune response and to supraphysiological levels of T-cell activation useful in the clinical setting (see Discussion Section).

In order to simplify the
mathematical analysis, the initial model was non-dimensionalized \cite{Itik2010}. The scaled resulting system of differential equations
is given by:%
\begin{equation}
\frac{dx_{1}}{dt}=x_{1}\left( 1-x_{1}\right)
-a_{12}x_{1}x_{2}-a_{13}x_{1}x_{3},  \label{x1}
\end{equation}%
\begin{equation}
\frac{dx_{2}}{dt}=r_{2}x_{2}\left( 1-x_{2}\right) -a_{21}x_{1}x_{2},
\label{x2}
\end{equation}%
\begin{equation}
\frac{dx_{3}}{dt}=\frac{r_{3}x_{1}x_{3}}{x_{1}+k_{3}}%
-a_{31}x_{1}x_{3}-d_{3}x_{3}.  \label{x3}
\end{equation}

The variables $x_{1}$, $x_{2}$ and $x_{3}$ denote, respectively, the population numbers of tumor cells, healthy cells and effector immune cells
against their specific maxima carrying capacities $k_{1}$, $k_{2}$ and $k_{3}$ (see Section 4 in \cite{Itik2010}). Parameter $a_{12}$ is the tumor cells inactivation rate by the healthy cells; 
 $a_{13}$ is the tumor cells inactivation rate by the effector cells; $r_{2}$ is the intrinsic growth rate of the
healthy tissue cells; $a_{21}$ is the healthy cells inactivation rate by the tumor cells; $r_3$ corresponds to the activation rate of effector cells due to tumor cells' antigen recognition; $a_{31}$ is the effector cells inactivation rate by the tumor cells. Finally,  $d_{3}$ is the density-dependent death rate of the effector cells (see \cite{Itik2010} for a detailed description of the model parameters). 

We want to notice that the inactivation rate (or the elimination rate) of tumor cells by the action of the effector immune cells (modeled with the last term in Eq. (1)) is assumed to be proportional to the number of effector immune cells, and no saturation is considered. A mechanism of elimination of tumor cells is given by the release of cytotoxic granules by the effector cells that impair or destroy tumor cells. Effector cells can clonally expand after antigen recognition, so the model assumes that they can be present in excess if needed. Hence, no saturation is considered for this term. The activation of effector immune cells due to antigen recognition used in the first term of Eq. (3) can be viewed as a Holling-II functional response, typically used to model predator feeding saturation in ecological dynamical systems. For our system, it is assumed a decelerating activation rate at increasing number of tumor cells since the activation of effector immune cells is limited by their requirement to recognize the tumor antigens in the context of the Antigen Presenting Cells (APCs). In this case, a process of cell-cell interaction and receptor recognition is required between APCs and tumor cells prior to activation, and thus an increasing number of tumor cells does not necessarily involve an increasing activation of effector cells.

The dynamics of this model is very rich, and both ordered (e.g., stable points or periodic orbits) and disordered (i.e., chaos) dynamics can be found for different parameter values \cite{Itik2010,Duarte2013}.

The model parameters will be fixed, if not otherwise specified, following \cite{Itik2010}, i.e., $a_{12}=1$; $a_{21}=1.5$; $d_{3}=0.5$; $k_{3}=1$; $r_{2}=0.6$;  $a_{13} = 2.5$; $a_{31} = 0.2$. This set of parameter values can involve chaos for a wide range of $r_3$ values (see below).

\section{Homotopy analysis method}

As in the cancer model explored in this article, many practical situations can be modeled with 
different types of systems of ordinary differential equations of the form%
\begin{equation}
\overset{.}{x}_{i}=f_{i}\left( t,x_{1},...,x_{n}\right) ,\text{ }%
x_{i}(t_{0})=x_{i,0},\text{ }i=1,2,...,n.  
\label{EDOs}
\end{equation}
Firstly, according to homotopy analysis method (HAM) \cite{Liao2003}, each equation of the system (%
\ref{EDOs}) is written in the form%
\begin{equation*}
N_{i}\left[ x_{1}(t),x_{2}(t)...,x_{n}(t)\right] =0,\text{ }i=1,2,...,n,
\end{equation*}%
where $N_{1},N_{2},...,N_{n}$ are nonlinear operators, $t$ denotes the
independent variable and $x_{1}(t),x_{2}(t)...,x_{n}(t)$ are the unknown
functions. From a generalization of the traditional homotopy method, Liao
has stablished in \cite{Liao2003} the so-called \textit{zeroth-order
deformation equation }%
\begin{equation}
\left( 1-q\right) L\left[ \phi _{i}\left( t;q\right) -x_{i,0}(t)\right]
=qh_{0}N_{i}\left[ \phi _{1}\left( t;q\right) ,...,\phi _{n}\left(
t;q\right) \right] ,  
\label{ZerothDeformation}
\end{equation}%
where $q\in \left[ 0,1\right] $ is an embedding parameter, $h_{0}$ is a
non-zero auxiliary artifitial parameter, $L$ is an auxiliary linear
operator, $x_{i,0}(t)$ are initial guesses and $\phi _{i}\left( t;q\right) $
are unknown functions. It is important to emphasize that, in the frame of
HAM, there is a great freedom to choose auxiliary entities such as $h_{0}$, $%
L$ and base functions for the representation of the solution $x_{i}(t)$.
Specifically, we can use in the construction of the solution $x_{i}(t)$ base
functions such as polynomials, exponentials, rational functions, etc. It is
obvious that when $q=0$ and $q=1$, both%
\begin{equation*}
\phi _{i}\left( t;0\right) =x_{i,0}(t)\text{ and }\phi _{i}\left( t;1\right)
=x_{i}(t)
\end{equation*}%
hold. According to (\ref{ZerothDeformation}), as $q$ increases from $0$ to $%
1 $, the function $\phi _{i}\left( t;q\right) $ varies from the initial
guess $x_{i,0}(t)$ to the solution $x_{i}(t)$. Expanding $\phi _{i}\left(
t;q\right) $ in Taylor series with respect to $q$, we obtain%
\begin{equation}
\phi _{i}\left( t;q\right) =x_{i,0}(t)+\sum\limits_{m=1}^{+\infty
}x_{i,m}(t)q^{m},  \label{Phitq}
\end{equation}%
where%
\begin{equation}
x_{i,m}(t)=\left. \frac{1}{m!}\frac{\partial ^{m}\phi _{i}\left( t;q\right) 
}{\partial q^{m}}\right\vert _{q=0}.  \label{mDeriv}
\end{equation}%
As stated by Liao \cite{Liao2003}, if the auxiliary linear operators, the
base functions and the auxiliary parameter $h_{0}$ are properly chosen, then
the series (\ref{Phitq}) converges at $q=1$ and%
\begin{equation*}
x_{i}(t)=\phi _{i}\left( t;1\right) =x_{i,0}(t)+\sum\limits_{m=1}^{+\infty
}x_{i,m}(t),
\end{equation*}%
which is one of the solutions of the original nonlinear equations. Taking
the $m^{th}$-order homotopy-derivative of the $zero^{th}$-order Eqs. (5), and using the corresponding properties, we have the $%
m^{th} $-order deformation equations%
\begin{widetext}
\begin{equation}
L\left[ x_{i,m}(t)-\chi _{m}x_{i,m-1}(t)\right] =h_{0}R_{i,m}\left[
x_{1,m-1}(t),...,x_{n,m-1}\left( t\right) \right] ,\text{ \ }i=1,2,...,n,
\label{mthOrderDeform}
\end{equation}%
\end{widetext}
where%
\begin{equation*}
R_{i,m}\left[ x_{1,m-1}(t),...,x_{n,m-1}\left( t\right) \right] =\left. 
\frac{1}{(m-1)!}\frac{\partial ^{m-1}N_{i}\left[ \phi _{1}\left( t;q\right)
,...,\phi _{n}\left( t;q\right) \right] }{\partial q^{m-1}}\right\vert _{q=0}
\end{equation*}%
and%
\begin{equation*}
\chi _{m}=\left\{ 
\begin{array}{c}
0,\text{ \ }m\leq 1 \\ 
\\ 
1,\text{ \ }m>1%
\end{array}%
\right. .
\end{equation*}%

It is important to notice that each function $x_{i,m}(t)$ $(m\geq 1)$ is
governed by the linear family of equations (\ref{mthOrderDeform}). This way
the HAM converts a complicated nonlinear problem into simpler linear
sub-problems. For some strongly nonlinear problems it is appropriate to use
the step homotopy analysis method (SHAM). This analytical technique is based
on a modification of the standard HAM, that we have just described, which
allows us to obtain a one-parameter family of explicit series solutions in a
sequence of intervals. For more details, we refer the reader to Ref. \cite{Alomari2010} (and references therein), where the SHAM is also explained in detail for the fractional Lorenz system. At this moment, after the previous considerations, we
are able to apply the HAM and the SHAM for solving analytically the
Itik-Banks cancer growth model.

Let us consider the Eqs. (\ref{x1})-(\ref{x3}) subject to the initial
conditions%
\begin{equation*}
x_{1}(0)=IC_{1},\text{ \ \ }x_{2}(0)=IC_{2},\text{ \ \ }x_{3}(0)=IC_{3}.
\end{equation*}%
Following the HAM, it is straightforward to choose%
\begin{equation*}
x_{1,0}(t)=IC_{1},\text{ \ \ }x_{2,0}(t)=IC_{2},\text{ \ \ }%
x_{3,0}(t)=IC_{3},
\end{equation*}%
as our initial approximations of $x_{1}(t),$ $x_{2}\left( t\right) $ and $%
x_{3}\left( t\right) $, respectively. In this work we will use $%
IC_{1}=0.13858...,$ $IC_{2}=0.69568...,$ $IC_{3}=0.01380...$. We choose the
auxiliary linear operators%
\begin{equation*}
L\left[ \phi _{i}\left( t;q\right) \right] =\frac{\partial \phi _{i}\left(
t;q\right) }{\partial t}+\phi _{i}\left( t;q\right) ,
\end{equation*}%
with the property $L\left[ C_{i}e^{-t}\right] =0$, where $C_{i}$ 
are integral constants (hereafter $i=1,2,3$). The Eqs. (\ref{x1})-(\ref{x3}) suggest the definition of the nonlinear operators $N_{1},$ $N_{2}$
and $N_{3}$ as%
\begin{eqnarray*}
N_{1}\left[ \phi _{1}\left( t;q\right) ,\phi _{2}\left( t;q\right) ,\phi
_{3}\left( t;q\right) \right] &=&\frac{\partial \phi _{1}\left( t;q\right) }{%
\partial t}-\phi _{1}\left( t;q\right) +\phi _{1}^{2}\left( t;q\right)
+a_{12}\phi _{1}\left( t;q\right) \phi _{2}\left( t;q\right) + \\
&&+a_{13}\phi _{1}\left( t;q\right) \phi _{3}\left( t;q\right) ,
\end{eqnarray*}%
\begin{equation*}
N_{2}\left[ \phi _{1}\left( t;q\right) ,\phi _{2}\left( t;q\right) ,\phi
_{3}\left( t;q\right) \right] =\frac{\partial \phi _{2}\left( t;q\right) }{%
\partial t}-r_{2}\phi _{2}\left( t;q\right) +r_{2}\phi _{2}^{2}\left(
t;q\right) +a_{21}\phi _{1}\left( t;q\right) \phi _{2}\left( t;q\right) ,
\end{equation*}%
\begin{eqnarray*}
N_{3}\left[ \phi _{1}\left( t;q\right) ,\phi _{2}\left( t;q\right) ,\phi
_{3}\left( t;q\right) \right] &=&\phi _{1}\left( t;q\right) \frac{\partial
\phi _{3}\left( t;q\right) }{\partial t}+k_{3}\frac{\partial \phi _{3}\left(
t;q\right) }{\partial t}-r_{3}\phi _{1}\left( t;q\right) \phi _{3}\left(
t;q\right) + \\
&&+a_{31}\phi _{1}^{2}\left( t;q\right) \phi _{3}\left( t;q\right)
+a_{31}k_{3}\phi _{1}\left( t;q\right) \phi _{3}\left( t;q\right) + \\
&&+d_{3}\phi _{1}\left( t;q\right) \phi _{3}\left( t;q\right)
+d_{3}k_{3}\phi _{3}\left( t;q\right) .
\end{eqnarray*}%
If $q\in \left[ 0,1\right] $ and $h_{0}$ the non-zero auxiliary parameter,
the $zero^{th}$-order deformation equations are of the following form%

\begin{equation}
\left( 1-q\right) L\left[ \phi _{i}\left( t;q\right) -x_{i,0}(t)\right]
=qh_{0}N_{i}\left[ \phi _{1}\left( t;q\right) ,\phi _{2}\left( t;q\right)
,\phi _{3}\left( t;q\right) \right] , 
\label{ZerothDeform1}
\end{equation}%
subject to the initial conditions%
\begin{equation*}
\phi _{1}\left( 0;q\right) =0.13858...,\phi _{2}\left( 0;q\right)
=0.69568...,\text{ }\phi _{3}\left( 0;q\right) =0.01380....
\end{equation*}%
For $q=0$ and $q=1$, the above $zero^{th}$-order equations (\ref{ZerothDeform1}%
) have the solutions%
\begin{equation}
\phi _{1}\left( t;0\right) =x_{1,0}(t),\text{ }\phi _{2}\left( t;0\right)
=x_{2,0}(t),\text{ }\phi _{3}\left( t;0\right) =x_{3,0}(t)
\label{SolPhiZero}
\end{equation}%
and%
\begin{equation}
\phi _{1}\left( t;1\right) =x_{1}(t),\text{ }\phi _{2}\left( t;1\right)
=x_{2}(t),\text{ }\phi _{3}\left( t;1\right) =x_{3}(t).  \label{SolPhiOne}
\end{equation}%
When $q$ increases from $0$ to $1$, the functions $\phi _{1}\left(
t;q\right) $, $\phi _{2}\left( t;q\right) $ and $\phi _{3}\left( t;q\right) $
vary from $x_{1,0}(t)$, $x_{2,0}(t)$ and $x_{3,0}(t)$ to $x_{1}(t)$, $%
x_{2}(t)$ and $x_{3}(t)$, respectively. Expanding $\phi _{1}\left(
t;q\right) $, $\phi _{2}\left( t;q\right) $ and $\phi _{3}\left( t;q\right) $
in Taylor series with respect to $q$, we have the homotopy-Maclaurin series%
\begin{equation}
\phi _{i}\left( t;q\right) =x_{i,0}(t)+\sum\limits_{m=1}^{+\infty}x_{i,m}(t)q^{m},  \label{MacLaurin1}
\end{equation}%
in which%
\begin{equation}
x_{i,m}(t)=\left. \frac{1}{m!}\frac{\partial^{m}\phi _{i}\left( t;q\right) 
}{\partial q^{m}}\right\vert _{q=0},
\label{X1mX2mX3m}
\end{equation}%
where $h_{0}$ is chosen in such a way that these series are convergent at $%
q=1$. Thus, through Eqs (\ref{SolPhiZero})-(\ref{X1mX2mX3m}), we have the
homotopy series solutions%
\begin{equation}
x_{i}(t)=x_{i,0}(t)+\sum\limits_{m=1}^{+\infty }x_{i,m}(t),
\end{equation}%
Taking the $m$th-order homotopy derivative of $zero$th-order Eqs. \eqref{ZerothDeform1}, and using the properties%
\begin{equation*}
D_{m}\left( \phi _{i}\right) =x_{i,m},\text{ }D_{m}\left( q^{k}\phi
_{i}\right) =D_{m-k}\left( \phi _{i}\right) =\left\{ 
\begin{array}{ll}
x_{i,m-k}, & \text{ \ \ }0\leq k\leq m \\ 
&  \\ 
0, & \text{ \ \ otherwise}%
\end{array}%
\right. ,
\end{equation*}%
\begin{equation*}
D_{m}\left( \phi _{i}^{2}\right) =\sum\limits_{k=0}^{m}x_{i,m-k}~x_{i,k},
\end{equation*}%
and
\begin{figure}[tbp]
\begin{center}
\includegraphics[width=0.425\textwidth]{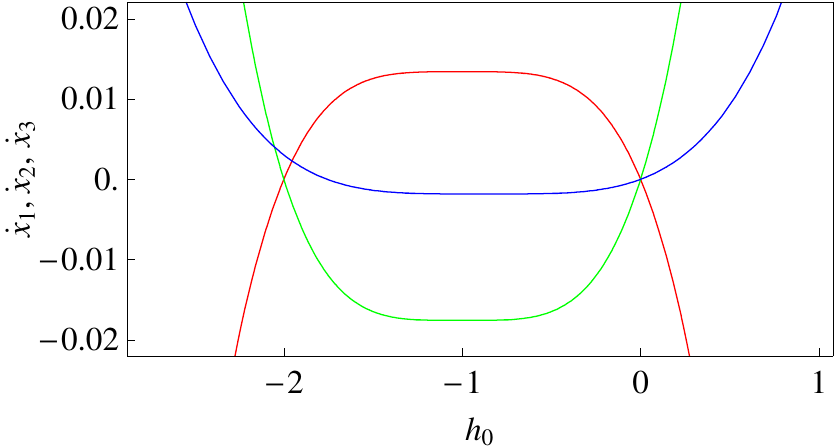}
\end{center}
\caption{Samples of $h_{0}$-curves for the Itik-Banks system under $12^{th}$%
-order approximation for $t=0$, $a_{13}=5.0$, $a_{31}=1.3$ and $r_{3}=4.5$
(red - $\protect\overset{\cdot }{x}_{1}\left( h_{0}\right) $, green - $%
\protect\overset{\cdot }{x}_{2}\left( h_{0}\right) $ and blue - $%
\protect\overset{\cdot }{x}_{3}\left( h_{0}\right) $). }
\label{h0Curves}
\end{figure}

\begin{equation*}
D_{m}\left( \phi _{i}\psi _{i}\right) =\sum\limits_{k=0}^{m}D_{k}\left( \phi
_{i}\right) ~D_{m-k}\left( \psi _{i}\right)
=\sum\limits_{k=0}^{m}x_{i,k}~y_{i,m-k},
\end{equation*}%
where $D_{m}$ means the $m$th-order derivative in order to $q$, we obtain the $m^{th}$-order deformation equations%
\begin{equation}
L\left[ x_{i,m}(t)-\chi _{m}x_{i,m-1}(t)\right] =h_{0}R_{i,m}\left[
x_{1,m-1}(t),x_{2,m-1}(t),x_{3,m-1}(t)\right] ,  \label{MDeform1}
\end{equation}%
with%
\begin{equation*}
\chi _{m}=\left\{ 
\begin{array}{c}
0,\text{ \ }m\leq 1 \\ 
\\ 
1,\text{ \ }m>1%
\end{array}%
\right.
\end{equation*}%
and the following initial conditions%
\begin{equation}
x_{1,m}(0)=0,\text{ }x_{2,m}(0)=0,\text{ }x_{3,m}(0)=0.  \label{IConditions}
\end{equation}
Defining the vector $\overrightarrow{x}_{m-1}=\left(
x_{1,m-1}(t),x_{2,m-1}(t),x_{3,m-1}(t)\right) ,$%
\begin{eqnarray}
R_{1,m}\left[ \overrightarrow{x}_{m-1}\right] &=&\overset{.}{x}%
_{1,m-1}(t)-x_{1,m-1}(t)+\sum\limits_{k=0}^{m-1}x_{1,m-1-k}(t)~x_{1,k}(t)+
\label{R1m1} \\
&&a_{12}\sum\limits_{k=0}^{m-1}x_{1,k}(t)~x_{2,m-1-k}(t)+a_{13}\sum%
\limits_{k=0}^{m-1}x_{1,k}(t)~x_{3,m-1-k}(t),  \notag
\end{eqnarray}
\begin{eqnarray}
R_{2,m}\left[ \overrightarrow{x}_{m-1}\right] &=&\overset{.}{x}%
_{2,m-1}(t)-r_{2}x_{2,m-1}(t)+r_{2}\sum%
\limits_{k=0}^{m-1}x_{2,m-1-k}(t)~x_{2,k}(t)+  \label{R1m2} \\
&&+a_{21}\sum\limits_{k=0}^{m-1}x_{1,k}(t)~x_{2,m-1-k}(t),  \notag
\end{eqnarray}%
and
\begin{eqnarray}
R_{3,m}\left[ \overrightarrow{x}_{m-1}\right] &=&\sum\limits_{k=0}^{m-1}%
\left( x_{1,k}(t)\overset{.}{x}_{3,m-1-k}(t)\right) +k_{3}\overset{.}{x}%
_{3,m-1}(t)-  \label{R2m3} \\
&&-r_{3}\sum\limits_{k=0}^{m-1}\left( x_{1,k}(t)~x_{3,m-1-k}(t)\right) + 
\notag \\
&&+a_{31}\sum\limits_{k=0}^{m-1}\left[ \left(
\sum\limits_{j=0}^{k}x_{1,k-j}(t)~x_{1,j}(t)\right) x_{3,m-1-k}\right] + 
\notag \\
&&+a_{31}k_{3}\sum\limits_{k=0}^{m-1}\left( x_{1,k}(t)~x_{3,m-1-k}(t)\right)
+d_{3}\sum\limits_{k=0}^{m-1}\left( x_{1,k}(t)~x_{3,m-1-k}(t)\right) + 
\notag \\
&&+d_{3}k_{3}x_{3,m-1}(t).  \notag
\end{eqnarray}
\begin{figure*}[tbp]
\begin{center}
\includegraphics[width=1.0\textwidth]{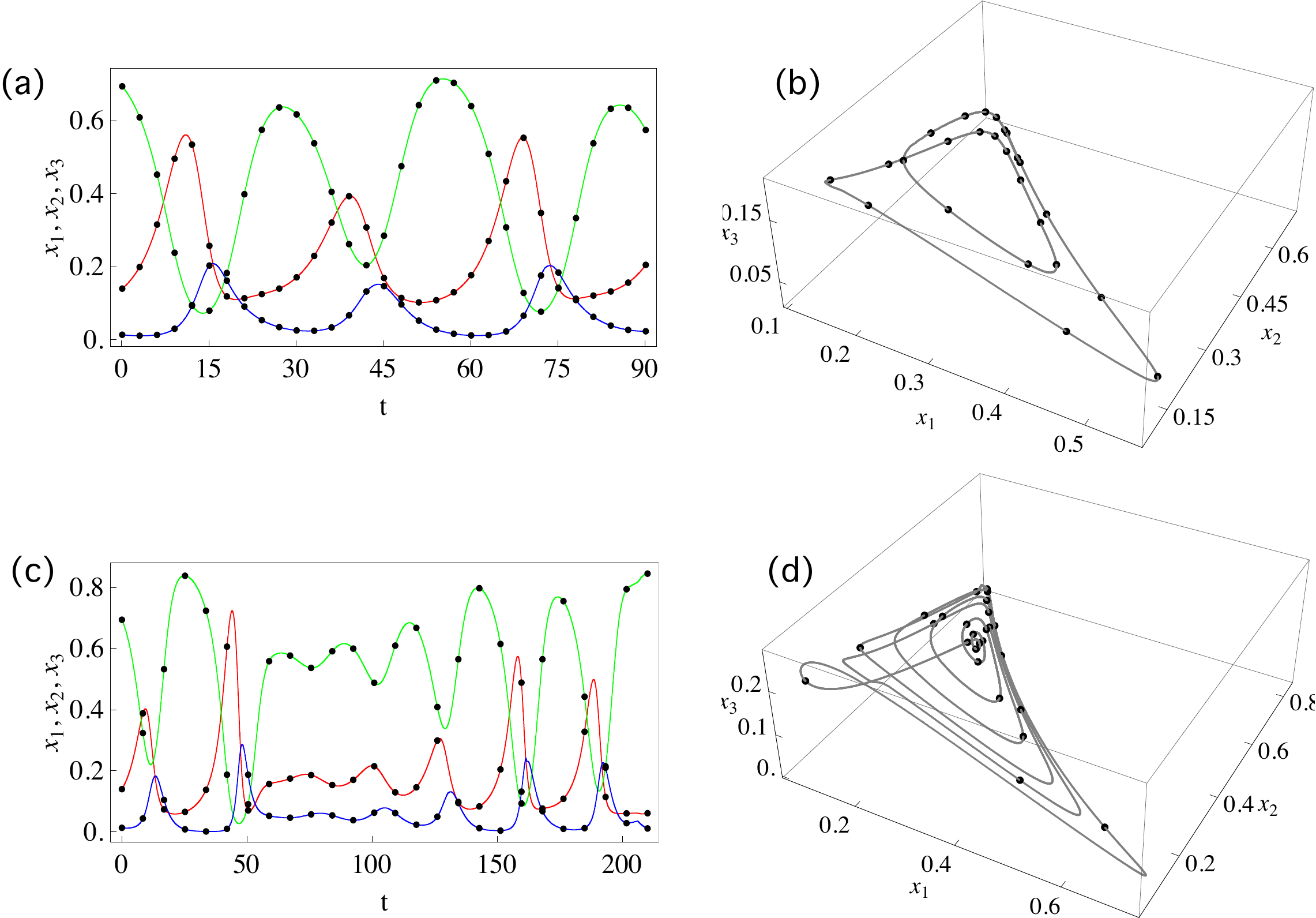}
\end{center}
\caption{Comparison between the homotopy solutions (solid lines) obtained with the \textit{Step Homotopy Analysis Method} (SHAM) developed in Section 3, and the
numerical simulations (black dots) for Eqs. \eqref{x1}-\eqref{x3}.  Time series of the dynamical variables [$x_{1}$ (red), $x_{2}$ (green) and $x_{3}$ (blue)], and the corresponding attractors, represented in the phase space ($x_{1}, x_{2}, x_{3}$). In (a) and (b) we display the period 2 dynamics, using $a_{13}=5$, $a_{31}=1.3$, and $r_{3}=4$. In (c) and (d) we show the chaotic attractor obtained setting $(a_{13},a_{31},r_{3})=(5,0.9435,4.5)$.}
\label{TSeriesAttractorSHAM}
\end{figure*}
Proceeding in this way, it is easy to solve the linear non-homogeneous Eqs. (15) at initial conditions (\ref{IConditions}) for all $m\geq
1 $, obtaining:
\begin{equation}
x_{i,m}(t)=\chi _{m}~x_{i,m-1}(t)+h_{0}~e^{-t}\int\limits_{0}^{t}e^{\tau
}R_{i,m}\left[ \overrightarrow{x}_{m-1}\right] d\tau ,  \label{SolX1m}
\end{equation}%
As an example, for $m=1$, we have:
\begin{eqnarray*}
x_{1,1}(t)
&=&0.13858-0.119376h_{0}+0.0964073a_{12}h_{0}+0.0019124a_{13}h_{0}+0.119376e^{-t}h_{0}-
\\
&&-0.0964073a_{12}e^{-t}h_{0}-0.0019124a_{13}e^{-t}h_{0}, \\
&& \\
x_{2,1}(t)
&=&0.69568+0.0964073a_{21}h_{0}-0.0964073a_{21}e^{-t}h_{0}-0.211709h_{0}r_{2}+0.211709e^{-t}h_{0}r_{2},
\\
&& \\
x_{3,1}(t)
&=&0.0138+0.000265021a_{31}h_{0}+0.0019124d_{3}h_{0}-0.000265021a_{31}e^{-t}h_{0}-
\\
&&-0.0019124d_{3}e^{-t}h_{0}+0.0019124a_{31}h_{0}k_{3}+0.0138d_{3}h_{0}k_{3}-
\\
&&-0.0019124a_{31}e^{-t}h_{0}k_{3}-0.0138d_{3}e^{-t}h_{0}k_{3}-0.0019124h_{0}r_{3}+
\\
&&+0.0019124e^{-t}h_{0}r_{3}.
\end{eqnarray*}%
It is straightforward to obtain terms for other values of $m$. In order to have an effective analytical approach of Eqs. (\ref{x1})-(\ref%
{x3}) for higher values of $t$, we use the step homotopy analysis method, in
a sequence of subintervals of time step $\Delta t$ and the $12^{th}$-order
HAM approximate solutions of the form:
\begin{equation}
x_{i}(t)=x_{i,0}(t)+\sum\limits_{m=1}^{11}x_{i,m}(t),  \phantom{x} {\rm{with}} \phantom{x} i = 1,2, 3.
 \label{ApproximX1}
\end{equation}
at each subinterval. With the purpose of determining the value of $h_{0}$ for each subinterval,
we plot the $h_{0}$-curves for Eqs. (\ref{x1})-(\ref{x3}) (see an example
for $t=0$ in Fig. \ref{h0Curves}).

Accordingly to SHAM, the initial values $x_{1,0}$, $x_{2,0}$ and $x_{3,0}$
will be changed at each subinterval, i.e., $x_{1}(t^{\ast })=IC_{1}^{\ast
}=x_{1,0}$, $x_{2}(t^{\ast })=IC_{2}^{\ast }=x_{2,0}$ and $x_{3}(t^{\ast
})=IC_{3}^{\ast }=x_{3,0}$ and we should satisfy the initial conditions $%
x_{1,m}(t^{\ast })=0$, $x_{2,m}(t^{\ast })=0$ and $x_{3,m}(t^{\ast })=0$ for
all $m\geq 1$. So, the terms $x_{1,1}$, $x_{2,1}$ and $x_{3,1},$ presented
before as an example for $m=1$, take the form:
\begin{eqnarray*}
x_{1,1}(t)
&=&0.13858-0.119376h_{0}+0.0964073a_{12}h_{0}+0.0019124a_{13}h_{0}+0.119376e^{-(t-t^{\ast })}h_{0}-
\\
&&-0.0964073a_{12}e^{-(t-t^{\ast })}h_{0}-0.0019124a_{13}e^{-(t-t^{\ast
})}h_{0}, \\
&& \\
x_{2,1}(t) &=&0.69568+0.0964073a_{21}h_{0}-0.0964073a_{21}e^{-(t-t^{\ast
})}h_{0}-0.211709h_{0}r_{2}+ \\
&&+0.211709e^{-(t-t^{\ast })}h_{0}r_{2}, \\
&& \\
x_{3,1}(t)
&=&0.0138+0.000265021a_{31}h_{0}+0.0019124d_{3}h_{0}-0.000265021a_{31}e^{-(t-t^{\ast })}h_{0}-
\\
&&-0.0019124d_{3}e^{-(t-t^{\ast
})}h_{0}+0.0019124a_{31}h_{0}k_{3}+0.0138d_{3}h_{0}k_{3}- \\
&&-0.0019124a_{31}e^{-(t-t^{\ast })}h_{0}k_{3}-0.0138d_{3}e^{-(t-t^{\ast
})}h_{0}k_{3}-0.0019124h_{0}r_{3}+ \\
&&+0.0019124e^{-(t-t^{\ast })}h_{0}r_{3}.
\end{eqnarray*}%
Identical changes occur naturally for the other terms. As a consequence, the
semi-analytical solutions are:
\begin{equation}
x_{i}(t)=x_{i}(t^{\ast })+\sum\limits_{m=1}^{11}x_{i,m}(t-t^{\ast }), \phantom{x} {\rm{with}} \phantom{x} i = 1,2,3.
\end{equation}
In general, we only have information about the values of $x_{1}(t^{\ast })$, 
$x_{2}(t^{\ast })$ and $x_{3}(t^{\ast })$ at $t^{\ast }=0$, but we can
obtain these values by assuming that the new initial conditions is given by
the solutions in the previous interval. Our previous calculations are in perfect agreement with numerical simulations (computed with an adaptive Runge-Kutta-Fehlberg method of order $7-8$).
In Fig. \ref{TSeriesAttractorSHAM}
we show the comparison of the SHAM analytical solutions and the numerical
solutions of the system under study, considering two dynamical regimes: period-2 dynamics (Fig. 2a and b) and chaos (Fig. 2c and d).

\section{Impact of effector immune cells activation in the dynamics}
The calculations developed in the previous section allow us to provide analytical approximations to the solutions of the
cancer model given by Eqs. (1-3). In this section we will use the homotopy solutions to explore the role of a key parameter of the model: the
stimulation and activation of the immune system cells (cytotoxic lymphocytes, CTLs) via the recognition of tumor cells antigens. This recognition process is parametrized in the model by means of $r_{3}$ and $k_{3}$.
We will here focus on parameter $r_{3}$, which can be interpreted as the
density-dependent activation rate of effector cells due to the recognition of the antigens present in the surface of tumor cells. 
The constant $k_{3}$ is a saturation parameter, and will be fixed following \cite{Itik2010}. By using the time trajectories
obtained from Eq. (21), we will first investigate the effect of increasing
the activation rate of effector cells in the deterministic dynamics. Then, we will add stochasticity to the homotopy solutions in order to explore the impact of demographic fluctuations in the overall dynamics of the system under investigation.
\begin{figure*}[tbp]
\begin{center}
\includegraphics[width=.893\textwidth]{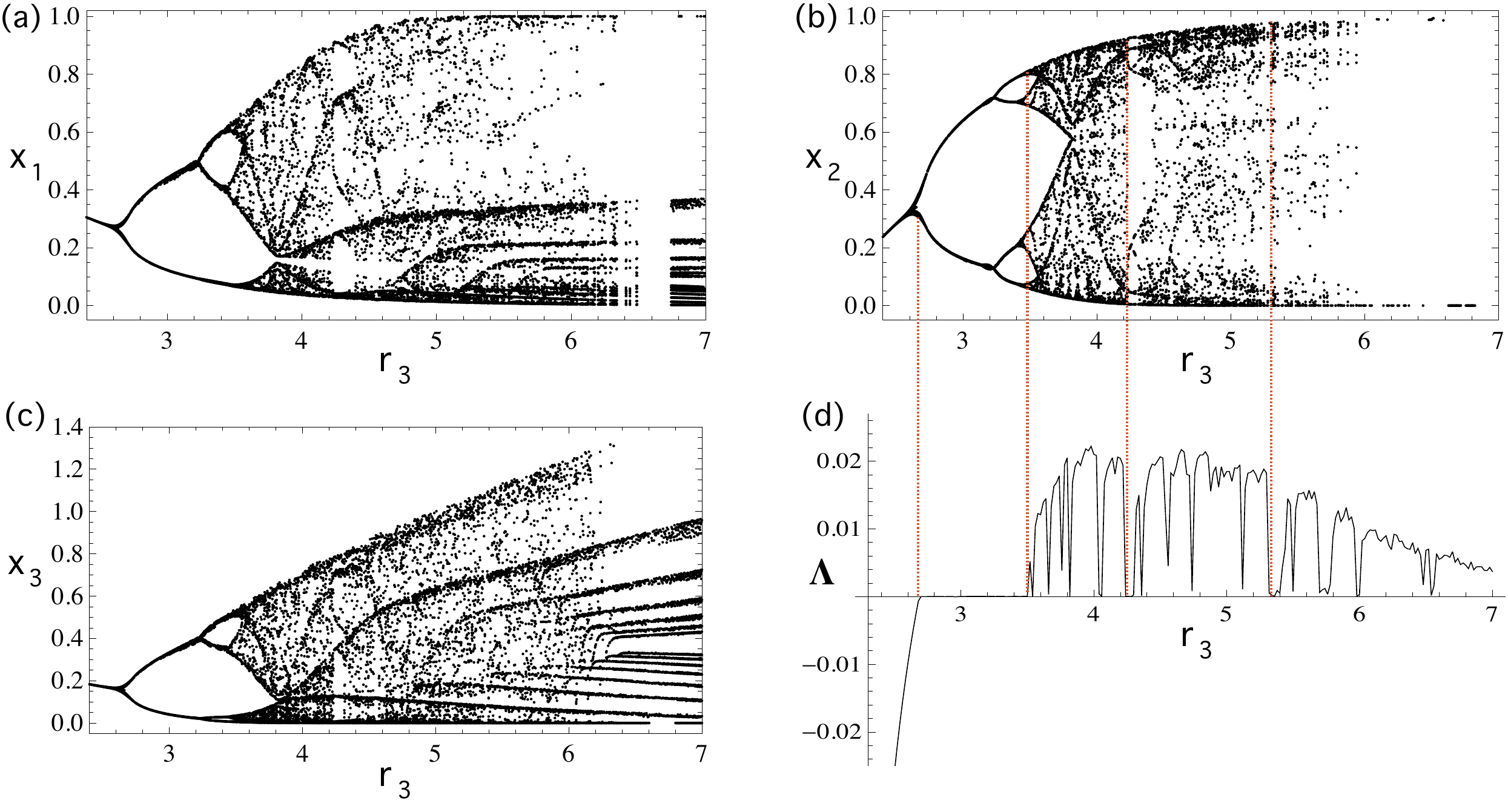}
\end{center}
\caption{Bifurcation diagrams obtained from the deterministic homotopy solutions, using the activation rate of efector immune cells (within the range $2.4 \leq r_{3}\leq 7$) as the control parameter. We plot the local maxima and minima of the homotopy solutions for the three model variables:  (a) tumor cells: $x_{1}$, (b) healthy cells: $x_{2}$, and (c) effector immune cells: $x_{3}$. Panel (d) shows the maximum Lyapunov exponent, $\Lambda$, for the same range $2.4 \leq r_{3}\leq 7$. The first red dashed line indicates a Hopf bifurcation, while the second one indicates the lowest value of $r_3$ where $\Lambda > 0$ i.e., chaos. Further increase of parameter $r_3$ involves other bifurcations e.g., $r_3 \sim 4.21$ or $r_3 \sim 5.35$.}
\label{BifCancer_r3}
\end{figure*}

The deterministic dynamics tuning $r_3$ are displayed in Fig. 3 by means of bifurcation diagrams built with the homotopy solutions. To build the bifurcation diagrams we computed a time series using the homotopy solutions for each value of $r_3$, and we recorded the local maxima and minima after discarding some transient. By using this approach it is shown that the increase of $r_3$ involves a period-doubling bifurcation scenario i.e., Feigenbaum cascade, that causes the entry of the cell populations into chaotic dynamics. For $r_3 \gtrsim 2.6$ the dynamics suffers the first bifurcation which switches the dynamics from a stable equilibrium towards a periodic orbit (dashed red line at the left in Fig. 3d). Further increase of $r_3$ involves period-doubling bifurcations, and, for $r_3 \gtrsim 3.5$ the dynamics undergo irregular fluctuations, which are confirmed to be chaotic with the computation of the maximal Lyapunov exponent, $\Lambda$ (Fig. 3d). $\Lambda$ has been computed within the range $2.4 \leq r_3 \leq 7$ from the model Eqs. (1-3) using a standard method \cite{Chua1989}. The bifurcation diagrams reveal that the population of cells undergoes larger fluctuations at increasing $r_3$, and populations can, at a given time point, be close to zero population values (extinction), as discussed for single-species chaotic dynamics \cite{Berryman1989}. That is, one might expect extinctions at increasing values of $r_3$.


In order to analyze extinction scenarios for the populations of cells in our model, we will use the homotopy solutions developed in Section 3, including a noise term simulating demographic stochasticity. Demographic stochasticity may play an important role at the initial stages of tumor progression, where the number of tumor cells is low compared to the population of healthy cells. Hence, we will assume that noise in tumor cells populations and in effector cells populations is larger than in healthy cells populations. Hence, we will include an additive stochastic term, $\xi_i(t)$, to the homotopy solutions, now given by:
\begin{equation}
x_{i}(t)=x_{i}(t^{\ast })+\sum\limits_{m=1}^{11}x_{i,m}(t-t^{\ast }) + \xi_i(t) \cdot (t - t^*), \phantom{x} {\rm{with}} \phantom{x} i = 1,2, 3.
\label{general_noise}
\end{equation}
Here $\xi_i(t)$ is a time-dependent random variable with uniform distribution  i.e., $\xi_{i=1,2,3}(t) \in U(-\sigma_i,\sigma_i)$ that simulates demographic fluctuations, where parameter $\sigma_i$ corresponds to the amplitude of the fluctuations. Previous works followed this approach to simulate decorrelating demographic noise in metapopulations \cite{Allen1993} and host-parasitoid \cite{Sardanyes2011} dynamics. Notice that the noise term is scaled by the time-step used to compute the homotopy solutions. As mentioned, in our model approach we will assume that the population of healthy cells is much larger than the populations of tumor and effector immune cells, setting $\sigma_2 = 0$. Hence, noise terms will be introduced to tumor and effector cells populations by means of $\sigma_{1,3} > \sigma_2 = 0$.  We notice that we can analyze the deterministic dynamics setting $\sigma_{1, 2, 3} = 0$.
Furthermore, the initial population numbers (initial conditions) for healthy cells populations are fixed to their carrying capacity $x_2(0)=k_2=1$, using $x_{1,3}(0) < x_2(0)$. 
  \begin{figure*}[tbp]
\begin{center}
\includegraphics[width=.9\textwidth]{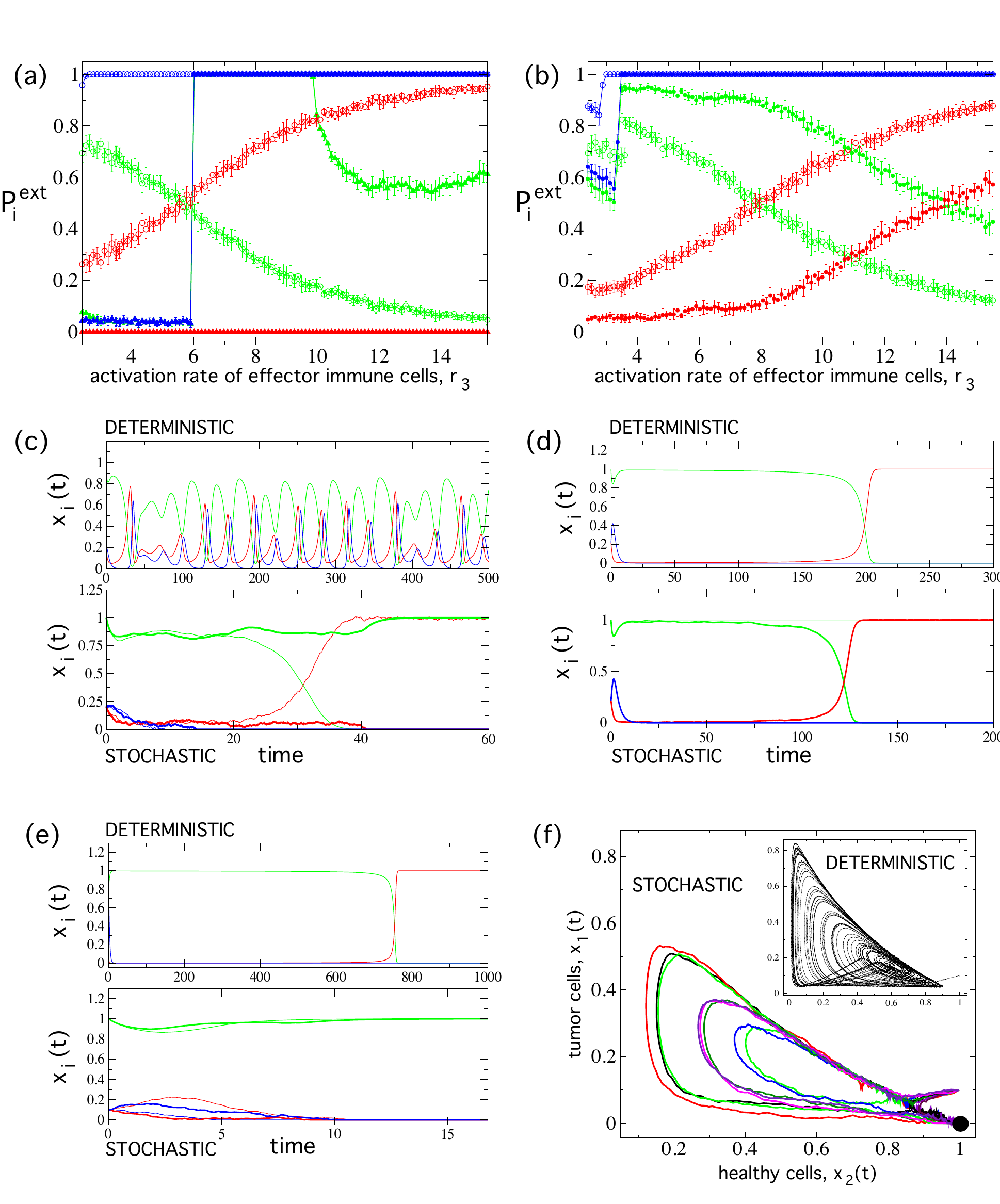}
\end{center}
\caption{Extinction probabilities, $P_i^{ext}$, of tumor ($i=1$, red), healthy ($i=2$, green), and effector immune ($i=3$, blue) cells at increasing activation rates of the immune cells (parameter $r_3$), fixing all other model parameters following \cite{Itik2010}. (a) Extinction probabilities for the deterministic dynamics (solid triangles), and for the stochastic dynamics using $\sigma_{1,3} = 0.1 > \sigma_2 = 0$ (open circles). (b) Extinction probabilities using $\sigma_{1,3} = 0.05$ (open circles) and $\sigma_{1,3} = 0.01$ (solid circles), both also with $\sigma_{2} = 0$. Each data point is the mean ($\pm SD)$ computed over $10$ replicas. Each of these replicas was obtained computing the extinction probability for each variable over $200$ time series of length $t=10^4$, starting from random low initial conditions for effector and tumor cells, $x_{1,3}(0) < x_2(0) = 1$ (see Section 4). In the lower panels we display the deterministic and stochastic dynamics for different values of $r_3$, with: $r_3 = 3.7$ (c); $r_3 = 9$ (d); and $r_3 = 12.5$. For the stochastic dynamics we display two different runs represented with thin and thick trajectories for each variable, with: (c) $\sigma_{1,3} = 0.1$; (d) $\sigma_{1,3} = 0.01$; and (e) $\sigma_{1,3} = 0.05$, all with $\sigma_2 = 0$. (f) Dynamics projected in the phase space $(x_2,x_1)$  using $r_3 = 4$. We display $10$ stochastic trajectories using $\sigma_{1,3} = 0.05$. The inset in (f) displays the chaotic attractor for the same initial conditions and parameter values with $\sigma_{1...3 = 0}$ (i.e., deterministic dynamics).}
\label{ext_chaos}
\end{figure*}

Using the homotopy solutions, we compute the extinction probabilities, $P_{i=1,2,3}^{ext}$, for each of the cell populations at increasing values of activation rates of effector immune cells, $r_3$. The extinction probabilities are computed as follows: for each value of $r_3$ analyzed, we built $200$ different time series with the homotopy solutions using $t = 10^4$. Over these $200$ time series, we calculated the number of time series for each variable fulfilling the extinction condition of variable $i$, assumed to occur when $x_i(t) \leq 10^{-30}$, normalizing the number of extinction events over $200$. Then, we repeated the same process $10$ times (replicas), and we computed the mean $(\pm SD)$ of the normalized extinction events over these $10$ replicas. Following the previous procedure, we consider random initial conditions for tumor and effector cells, setting $x_2(0) = 1$ and $x_{1,3}(0) < x_2(0)$, instead of using a single initial condition for each variable for all time series. Specifically, we will consider random initial populations of tumor and effector cells following a uniform distribution within the range $(0, 0.2]$.
The results are displayed in Fig. 4 using parameter values from \cite{Itik2010}, except for the tuned parameter $r_3$. The deterministic simulations (solid triangles in Fig. 4a) reveal that extinction probabilities for tumor cells is zero within the range analyzed i.e., $2.4 \leq r_3 \leq 15.5$. For low values of $r_3$, the extinction probabilities for healthy and effector cells remain close and low ($P_{2,3}^{ext} \sim 0.05$). Beyond $r_3 \gtrsim 6$, $P_{2,3}^{ext}$ drastically increases and extinctions take place with probability $1$. Such extinction value is maintained for effector cells at increasing $r_3$. The extinction probability of healthy cells diminishes beyond $r_3 \gtrsim 9.8$ to $P_2^{ext} \sim 0.6$. Counterintuitively, these results indicate that  increasing activation of effector immune cells (using the parameter values from \cite{Itik2010}) does not involve tumor cells extinction or low extinction probabilities for healthy cells, due to the complexity of the dynamics in the chaotic or fluctuating regimes.

Now, we focus on the effect of demographic stochasticity in the overall dynamics of the system. Figure 4a displays the same analyses performed with the deterministic approach, but now using $\sigma_{1,3} = 0.1$ (recall $\sigma_2 = 0$). The observed extinctions patterns drastically change. For instance, the extinction probability of tumor cells, $P_{1}^{ext}$, ranges from $P_{1}^{ext} \sim 0.3$ to $P_{1}^{ext} \sim 0.97$ within the range $2.4 \leq r_3 \leq 15.5$. Moreover, the extinction probability of healthy cells significantly decreases at increasing $r_3$, having values of $P_{1}^{ext} \sim 0.05$ for $r_3 \gtrsim 12.2$. These results clearly indicate that when demographic noise is high (e.g., at initial tumor progression stages) stochastic fluctuations can involve increasing extinction probabilities of tumor cells and increasing survival probabilities of healthy cells when $r_3$ grows. In the stochastic simulations, effector immune cells always reached extinction, except for the cases with small $r_3$ and low noise amplitudes (Fig. 3b). Similar results were obtained by using $\sigma_{1,3} =0.05$ (open circles in Fig. 4b) and $\sigma_{1,3} =0.01$ (solid circles in Fig. 4b). As expected, the decrease of the noise levels involves lower extinction probabilities for both tumor and healthy cells, although the same tendencies are preserved i.e., increasing $r_3$ enlarges tumor cells extinctions and decreases host cells extinctions. 

Figure 4(c-d) displays several time series for different values of $r_3$ and noise intensities, also representing the deterministic dynamics. For instance, in Fig. 4c (setting $r_3 = 3.7$) the deterministic dynamics is chaotic and no extinctions are found. However, the stochastic dynamics (here using $\sigma_{1,3} = 0.1$) can cause extinction or survival of tumor and healthy cells also with $r_3 = 3.7$. In Fig. 4d we use $r_3 = 9$. The deterministic dynamics for this case involves outcompetition of healthy and effector cells by tumor cells. However, the stochastic dynamics ($\sigma_{1,3} = 0.01$) can involve either the extinction or survival of tumor cells. Finally, Fig. 4e displays the dynamics using $r_3 = 12.5$. For this case, the deterministic dynamics also involves dominance of tumor cells, but the stochastic dynamics (with $\sigma_{1,3} = 0.05$) involves an extinction probability of tumor cells of $P_1^{ext} \sim 0.8$. In Fig. 4f we display the trajectories projected in the phase space $(x_2(t), x_1(t))$ using $r_3 = 4$ and $\sigma_{1,3} = 0.05$. The main plot shows ten stochastic trajectories that reach the $(0,1,0)$ attractor (solid black circle) that involves the survival of healthy cells and the extinction of both effector and tumor cells. The inset displays the deterministic dynamics also for $r_3 = 4$, which is governed by the chaotic attractor.

Finally, we want to note that the same qualitative extinction patterns were obtained using parameter values explored in \cite{Duarte2013}, setting: $a_{31} = 0.9435$ and $a_{13} = 5$. Moreover, all the previous simulations (using parameter values from Refs. \cite{Itik2010} and \cite{Duarte2013}) were repeated using different extinction thresholds i.e., $x_i(t) \leq 10^{-10}$ and $x_i(t) \leq 10^{-20}$, and the extinction probabilities remained qualitatively equal for both deterministic and stochastic dynamics (results not shown).

\section{Discussion}
In this article, a semi-analytic method to find approximate solutions for nonlinear differential equations -
the step homotopy analysis method (SHAM) - is applied to solve a cancer nonlinear model initially proposed by Itik and Banks \cite{Itik2010}. With this algorithm,
based on a modification of the homotopy analysis method (HAM) proposed by
Liao \cite{Liao1992,Liao2003,Liao2007}, three coupled nonlinear differential equations are replaced
by an infinite number of linear subproblems. This modified method has the
advantage of giving continuous solutions within each time interval, which is
not possible by purely numerical techniques. Associated to the explicit
series solutions there is an auxiliary parameter, called convergence-control
parameter, that represents a convenient way of
controlling the convergence of approximation series, which is a critical
qualitative difference in the analysis between HAM/SHAM and other methods.

The model by Itik and Banks \cite{Itik2010} considers the dynamics of three interacting cell types: healthy cells, tumor cells, and effector immune cells (i.e., CD8 T cells, also named cytotoxic lymphocytes, CTLs). 
Our analytical results are found to be in excellent agreement with the numerical simulations. To the best of our knowledge, such
kind of explicit series solutions, corresponding to each of the dynamical
variables, have never been reported for the Itik-Banks cancer model. The results presented in this
article suggest that SHAM is readily applicable to more complex chaotic
systems such as Volterra-Lotka type models applied to cancer dynamics. In this work we used the homotopy solutions to investigate the impact of a key parameter in the dynamics of tumor growth: the activation of effector immune cells due to recognition of tumor antigens (parameter $r_3$).  Previous research has focused on other key parameters of the model by Itik and Banks. For instance, the active suppression of the immune response by the tumor cells has been recently explored in Ref. \cite{Duarte2013}. Interestingly, the dynamics were shown to be very sensitive to the suppression of the immune cells, involving an inverse period-doubling bifurcation scenario at increasing the suppression rate of immune cells \cite{Duarte2013}. For this case, strong chaos and low predictability was found at small suppression rates, and the chaotic dynamics became more predictable at increasing suppression values. The selective shutdown of the antitumor  immune response can also be achieved by the escape of the recognition of the cancer cells by the immune system by selection of non-immunogenic tumor cell variants and in influencing immune cells with a negative regulatory function,  such as regulatory T  cells and myeloid-derived suppressor cells Thus, the cell-killing activity of the cytotoxic CD8 T cells can be inhibited by the presence within the tumor tissue of immunosuppressive CD4+ regulatory T cells ($T_{reg}$  cells). The function of the $T_{reg}$s is essential for inducing tolerance to "self " antigens, preventing autoimmune reactions and for the downregulation of the immune response after the elimination of the antigenic source (such as pathogens, allogenic cells or cancer cells). However, their capacity to inhibit the innate and adaptive anti-tumor immune response also constitute a major obstacle to cancer immunotherapy.

We have used the homotopy solutions to characterize changes in the dynamics at increasing activation rates of the immune cells. Such a parameter is especially important since several clinical therapies are currently available to boost immune responses (see next paragraph). The increase of the immune cells activation rate is shown to cause a period-doubling bifurcation scenario that makes the system to enter into chaotic dynamics. Interestingly, the populations of tumor cells, although undergoing large fluctuations, are able to survive for all the range of $r_3$ analyzed. In order to simulate demographic stochasticity that might be found at early stages of tumorigenesis, we added noise terms to the homotopy solutions for tumor and effector immune cells populations. As a difference from the deterministic dynamics, we found that an increase of $r_3$ increases the extinction probabilities for tumor cells, also diminishing the extinction probabilities of healthy cells. These results suggest that possible therapies enhancing the activation of effector immune cells (see next paragraph) at early stages of tumor progression could result in higher probabilities of stochastic tumor clearance. lt is worth to note that the model proposed by Itik and Banks does not explicitly model the clonal expansion of immune cells after tumor antigen recognition that could make the noise in CTLs populations to be even smaller or negligible. It is known that after being activated, the population of CTLs is expanded in order to exert strong cytotoxic effects. Then, the CD8 response is downregulated by programmed cell death mechanisms to avoid over-activation of the immune system (Raval et al., 2014). Due to the complexity of the dynamics found at increasing $r_3$, it is not clear if clonal expansion would favor the extinction of tumor cells, as we would expect. In this sense, the effect of immune system activation together with production of large populations of effector immune cells due to clonal expansion (burst in the population of effector immune cells) should be modeled to determine if our observed results remain the same or change the probabilities of tumor cells extinction in response to increases in $r_3$.

Our results could be clinically relevant since several therapies to stimulate and activate immune cells are currently available. A foundational property of the immune system is its capacity to distinguish between the "self " and "non-self " antigens. In the context of an evolving tumor, it is likely that the tumoral cells will present to the immune cells a number of new antigens product of the genetic aberrations present in their genome. This mechanism is probably involved in the control of early tumors. However, it is known that cancer cells escape innate and adaptive immune responses by selection of non-immunogenic tumor cell variants (immunoediting) or by active suppression of the immune response (immunosubversion) (see \cite{Raval2014} for a review). Tumor antigens often elicit poor adaptive immune responses because they are recognized as "self-antigens" that induce tolerance, the natural mechanism of the body to prevent autoimmunity. The enhancement of the antitumor T cell responses by triggering TCR costimulatory molecules to break tolerance has been envisaged as a way to potentiate the antitumor immune functions. Agonists of the costimulatory tumor necrosis factor receptor (TNFR) family members, which include proteins involved in B and T cell development, survival, and immune activation, have been proven to enhance the antitumor immune responses. Preclinical and early clinical data of the use of agonists of 4-1BB (CD137) or OX40 (CD134) support further studies of these costimulatory molecules as potentiators of the antitumor response \cite{Schaer2014}. An increasingly successful anticancer strategy that aims to boost immune responses against tumor cells consists in enhancing the cell-killing activity of the cytotoxic CD8 T cells by the use of antibodies that block negative regulators of T-cell activation ("checkpoint inhibitors"). Fully humanized monoclonal antibodies blocking the inhibitory molecules Cytotoxic T-Lymphocyte antigen 4 (CTLL4, Ipilimumab, Tremelimumab) or Programmed Death Receptor-1 (PD-1, Nivolumab, MK-3475) have been proven to be useful in solid tumors such as melanoma, renal cell carcinoma, non small cell lung cancer or colorectal cancer (reviewed in \cite{Kyi2014}). More recently, the p110$\delta$ isoform of phosphoinositide-3-OH kinase (PI(3)K) activity has been shown to be required for the proliferation and differentiation of suppressive $T_{reg}$ cells induced by tumor cells. PI(3)K $\delta$ inhibitors have been proven to be able to preferentially inhibit CD4 $T_{reg}$ cells over effector CTLs, opening new ways to unleash the power of dormant anti-tumor immune cells \cite{Ali2014}. More recent and novel approaches suggest the possibility to increase CTLs activation by means of artificial APCs (see \cite{Eggermont2014} for further details).

Summarizing, our results suggest that potential therapies increasing activation rates of effector immune cells might be much more effective at early stages of tumor progression, when demographic noise becomes important in tumor cells populations. Our results also suggest that the stimulation of immune cells may not facilitate tumor clearance in cancers with large population numbers of tumor cells, as the deterministic approach is considering. Further research should also analyze the robustness and generality of our results to changes in the other model parameters. As discussed in \cite{Duarte2013}, it would be also interesting to explore the effect of increasing the activation of effector immune cells in solid tumors by means of a spatial version of the cancer model analyzed in this article.

\begin{acknowledgments}
We want to thank Jos\'e Aramburu for helpful comments on immunology and cancer dynamics. We also acknowledge the Department of Applied Mathematics and Analysis from Universitat de Barcelona for kindly providing us with the RKF-78 method used for numerical integration. This work was partially funded by the Bot\'in Foundation (JS), by ISCIII-grant PI13/00864 (GG-G), and by FCT/Portugal through project PEst-OE/EEI/LA0009/2013 (NM, JD).eloped (grant NSF PHY05-51164).
\end{acknowledgments}

\end{document}